\documentclass[aps,prb,twocolumn,superscriptaddress,longbibliography]{revtex4-2}
\usepackage{amsmath,amssymb}
\usepackage[pdftex]{hyperref,graphicx}
\hypersetup{colorlinks = true, urlcolor = blue, linkcolor = blue, citecolor = blue}
\usepackage{xcolor}
\usepackage{bm}
\usepackage {verbatim}

\usepackage[etex=true,export]{adjustbox}
\usepackage{makecell}
\usepackage{multirow}
\usepackage{tabularx,multirow,array,diagbox}
\usepackage{adjustbox}
\usepackage{booktabs}

\newcommand{\beq}{\begin{equation}}
\newcommand{\eeq}{\end{equation}}
\newcommand{\beql}{\begin{equation*}}
\newcommand{\eeql}{\end{equation*}}
\newcommand{\beqn}{\begin{eqnarray}}
\newcommand{\eeqn}{\end{eqnarray}}

\begin{document}
\title{ Generalization of Benalcazar-Bernevig-Hughes model to arbitrary dimensions}
\author{Xun-Jiang Luo}
%\email{luoxunjiang@whu.edu.cn}
\affiliation{School of Physics and Technology, Wuhan University, Wuhan 430072, China}
\author{Fengcheng Wu}
\email{wufcheng@whu.edu.cn}
\affiliation{School of Physics and Technology, Wuhan University, Wuhan 430072, China}
\affiliation{Wuhan Institute of Quantum Technology, Wuhan 430206, China}
%\date{\today}

\begin{abstract}
The Benalcazar-Bernevig-Hughes (BBH) model [Science $\bold{357}$, 61 (2017)], featuring bulk quadrupole moment, edge dipole moments, and corner states, is a paradigm of both higher-order topological insulators and topological multipole insulators. In this work, we generalize the BBH model to arbitrary dimensions by utilizing the Clifford algebra. For the generalized BBH model, the analytical solution of corner states can be directly constructed in a unified way. Based on the solution of corner states and chiral symmetry analysis, we develop a general boundary projection method to extract the boundary Hamiltonians, which turns out to be the BBH models of lower dimension and reveals the dimensional hierarchy.

%thoroughly reveal the boundary-localized multipole moments of lower dimension, indicating the existence of bulk multipole moment. 

\end{abstract}
\maketitle

\section{Introduction}
The past few years have witnessed the rapid development of higher-order topological phases (HOTPs) \cite{Benalcazar2017a,Benalcazar2017,Liu2017,Song2017,Langbehn2017,Khalaf2018,Geier2018,Ezawa2018,Schindler2018a,
Schindler2018,Xie2018,Wang2018a,Zhu2018,Queiroz2019,Yan2019,
Zhang2019b,Luo2019,Xie2019,Yue2019,
Okugawa2019,Liu2019,Zhang2019d,Trifunovic2019,Peng2020,Cerjan2020,
Chen2020a,Hua2020,Hu2020,Huang2020,Choi2020,Zhang2020a,Aggarwal2021,Khalaf2021,
Wang2021,Zhang2021a,Benalcazar2022,Tan2022,Lei2022,
Jia2022}, which include insulators \cite{Ezawa2018a,Park2019,Sheng2019,Chen2020,Zhang2020b,Ren2020,Liu2021}, superconductors \cite{Yan2018,Wang2018,Zhang2018,Zhu2019,Zhang2019,Volpez2019,Pan2019,Wu2020,Wu2020a,
Zhang2021,Chen2021a,Luo2021a,SD2022,Li12022}, and semi-metals \cite{Lin2018,Wang2020,Wang2022}. In a $d$-dimensional ($d$D) system, an $n$th-order topological phase features robust gapless states at its $(d-n)$D boundaries. For example, the second- and third-order topological phases in 3D host robust hinge states and corner states, respectively. The corner states in a system with chiral or particle-hole symmetry stay at exact zero energy, which can give rise to fractional corner charges in insulators \cite{Benalcazar2019,Watanabe2021,Jung2021} and non-Abelian anyons in superconductors \cite{Zhang2020c,Zhang2020d,Pahomi2020,Pan2022}. These exotic topological properties of HOTPs attract much attention.

A typical example of HOTPs is the Benalcazar-Bernevig-Hughes (BBH) model \cite{Benalcazar2017a,Benalcazar2017}, which features corner states and bulk quadrupole moment (a quantity that generalizes the electric dipole moment to 2D) \cite{Benalcazar2017a,Benalcazar2017}.  The 3D and 4D  counterparts of the 2D BBH model, which host bulk octupole and hexadecapole moments, respectively, have also been studied \cite{Benalcazar2017a,Benalcazar2017,Chen2019b}. Particularly, the 2D, 3D, and 4D BBH models have been experimentally realized in artificial systems, including phononic crystals \cite{Serra-Garcia2018,Mittal2019,Schulz2022}, acoustic crystals \cite{Qi2020,Xue2020}, and electrical circuits \cite{Imhof2018,Serra-Garcia2019,Bao2019,Zhang2020e}.  An interesting question is how to generalize the BBH model to arbitrary dimensions.

\begin{figure}
\centering
\includegraphics[width=3.2in]{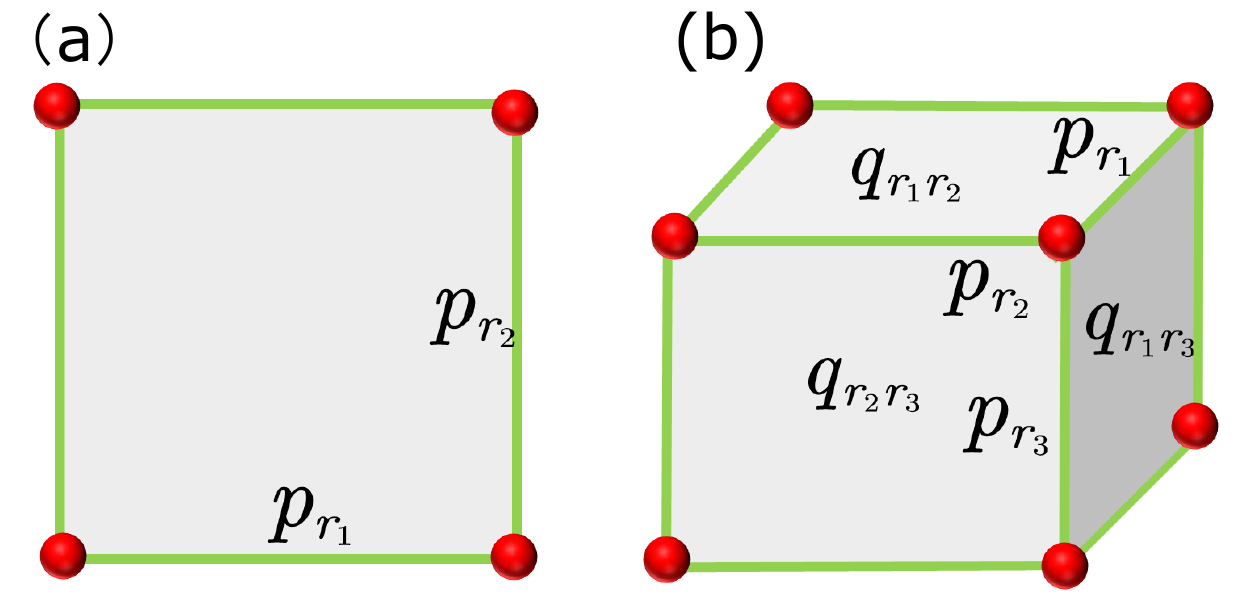}
\caption{(a) Schematic illustration for the edge dipole moments $(p_{r_1},p_{r_2})$ of topological quadrupole insulators. (b) Schematic illustration for the surface quadrupole moments $(q_{r_1r_2},q_{r_1r_3},q_{r_2r_3})$ and hinge dipole moments $(p_{r_1},p_{r_2},p_{r_3})$ of topological octupole insulators. }
\label{23D}
\end{figure}

The BBH model, as a kind of higher-dimensional generalization of the 1D Su-Schrieﬀer-Heeger (SSH) model \cite{Su1979}, exhibits dimensional hierarchy. The bulk quadrupole (octupole) moment of the 2D (3D) BBH model manifest themselves by the presence of edge dipole (surface quadrupole and hinge dipole moments), as schematically shown in Fig.~\ref{23D}(a) (Fig.~\ref{23D}(b)). In the seminal papers of Refs.~ \onlinecite{Benalcazar2017a,Benalcazar2017}, the nested Wilson loop method was developed to reveal boundary-localized multipole moment by exploring the Wannier band topology. However, this method increases its computational complexity sharply with the increasing of the dimension of the system. An alternative way of revealing the boundary localized multipole moments of the BBH model is to directly extract the boundary Hamiltonians of the system. For instance, the edge dipole moments in the 2D BBH model can be unveiled by its edge Hamiltonians, which turn out to be a two-band SSH model hosting dipole moment \cite{Li2020a,Luo2022}. Similarly, the  surface quadrupole moments and hinge dipole moments  of the 3D BBH model can be revealed by the surface and hinge Hamiltonians of the system, respectively. Therefore, the key point of uncovering the dimensional hierarchy of the BBH model is to analytically derive the boundary Hamiltonians through boundary projection in a systematic and generalized way.

In this work, we generalize the BBH model to arbitrary dimensions and present a unified Hamiltonian form, which is the summation of the extended Su-Schrieﬀer-Heeger (SSH) model (multi copies of the two-band SSH model) along different directions. Based on the thorough investigation of the 1D extended SSH model enabled by chiral symmetry analysis, we directly construct the analytical solution of corner states in a unified way. Based on the solution of corner states, we further develop a general boundary projection method to derive the boundary Hamiltonians for an arbitrary dimensional system, which turns out to be the BBH model of lower dimension and reveals the dimensional hierarchy.

%thoroughly unveils the boundary-localized multipole moments of the lower dimension and indicates the bulk multipole moment.
% analytical solution of the end states of the extended SSH model and chiral symmetries analysis, we directly construct the analytical solution of corner states, which possess a unified and elegant form. 

The rest of this paper is organized as follows. In Sec.~\ref{II}, we introduce the 1D extended SSH model. In Sec.~\ref{III}, we generalize the BBH model to arbitrary dimensions and construct the analytical solution of corner states.
In Sec.~\ref{IV}, we perform boundary projection analysis and extract the boundary Hamiltonians. In Sec.~\ref{V}, we conclude with a discussion and summary. Appendices \ref{Appendix I}-\ref{Appendix III} complement the main text with additional technical details.

\section{ 1D Extended SSH model}
\label{II}
We start our discussion by introducing the 1D extended SSH model, which will be used as a building block to construct the Hamiltonians of the BBH model in arbitrary dimensions.
The Bloch Hamiltonian of the 1D extended SSH model is
\beqn
h(k)=(t+\lambda\cos k)\gamma_{a}^{(d)}+\lambda\sin k\gamma_{b}^{(d)},
\label{II1}
\eeqn
where $t$ and $\lambda$ are model parameters and $k$ denotes the momentum. In Eq.~\eqref{II1}, $\gamma_{a}^{(d)}$ and $\gamma_{b}^{(d)}$ are $2^d\times 2^d$ anti-commuting Gamma matrices (see Appendix~\ref{Appendix I}),
\beqn
\{\gamma_{a}^{(d)},\gamma_{b}^{(d)}\}=0,\quad (\gamma_{a}^{(d)})^2=(\gamma_{b}^{(d)})^2=1.
\eeqn
The chiral symmetry of $h$ is explicitly defined by $C=i\gamma_{a}^{(d)}\gamma_{b}^{(d)}$ which satisfies $\{C,h\}=0$. Therefore, 1D Hamiltonian $h$ belongs to the AIII symmetry class and has a $Z$ topological classification \cite{Schnyder2008,Ryu2010,Chiu2016}.

To clearly display the topology of $h$, which does not depend on the representation of matrices $\gamma_{a}^{(d)}$ and $\gamma_{b}^{(d)}$, without loss of generality, we choose
\beqn
\gamma_{a}^{(d)}=\sigma_2\otimes\underbrace{\sigma_0\cdots\sigma_0}_{d-1},\quad\gamma_{b}^{(d)}=\sigma_1\otimes\underbrace{\sigma_0\cdots\sigma_0}_{d-1},
\label{II2}
\eeqn
where $\sigma_{1,2}$ are Pauli matrices and $\sigma_0$ is a $2\times 2$ identify matrix. Following Eq.~\eqref{II2}, $h$ is composed of $2^{d-1}$ blocks and each block behaves as a two-band SSH model. When $|t|<|\lambda|$, $h$ is topologically nontrivial and characterized by the winding number $\nu=2^{d-1}$, as explicitly shown in Appendix~\ref{Appendix II}.

In the open boundary condition, $h$ hosts $2^{d-1}$ end zero-energy states (EZESs) at each end owing to the bulk-boundary correspondence. The EZESs solution of $h$ can be analytically derived by directly solving the differential equation $h(r)|X_{\alpha}(r)\rangle=0$ (see Appendix~\ref{Appendix II}). Here,  the low-energy Hamiltonian $h(r)$ is obtained by expanding $h(k)$ at $k=0$ to the second order of $ k$ and replacing $k=-i\partial_r$, where $r$ is the 1D position coordinate. The EZESs are the eigenstates of chiral symmetry
\beqn
C=i\gamma_{a}^{(d)}\gamma_{b}^{(d)}=\sigma_z\otimes\underbrace{\sigma_0\cdots\sigma_0}_{d-1},
\eeqn
which has the eigenvalue $z=\pm 1$. The EZESs labeled by $z=-1$ and $z=1$ are localized close to end $r=0$ and $r=L$, respectively, where $L$ is the length of the 1D system. Therefore, the wave function of the 1D EZESs takes the form
\beqn
|X_{z}(r)\rangle=f_{z}(r)|\psi_{z}\rangle,
\label{II3}
\eeqn
where $2^d$-component spinor $|\psi_{z}\rangle$ satisfies $C|\psi_{z}\rangle=z|\psi_{z}\rangle$ and $f_{+}(r)$ and $f_{-}(r)$ are the real space localized wave function determined by parameters $t$ and $\lambda$.

We note that the topology of $h$ does not depend on the given representations of $\gamma_{a}^{(d)}$ and $\gamma_{b}^{(d)}$. In Appendix~\ref{Appendix II}, we calculate the winding number and EZESs wave function of $h$ for a general case.

\section{Hamiltonians and corner states}
\label{III}
\subsection{Model Hamiltonians}
Lower-dimensional topological states can be used to construct topological phases in higher dimensions \cite{Teo2010,Bomantara2019,Zhang2020f}. Here, we choose the 1D extended SSH model as the building block and construct the $d$D Hamiltonians
\beqn
\begin{aligned}
&\mathcal{H}_d(\bm k_d)=\sum_{s=1}^{d}h_s(k_s),\\
&h_s(k_s)=M_s(k_s)\gamma_{sa}^{(d)}+\lambda_s\sin k_s\gamma_{sb}^{(d)},
\label{III1}
\end{aligned}
\eeqn
where $\bm k_d=(k_1,\cdots, k_d)$ is the $d$D momentum vector, $s$ labels the different directions, $M_s(k_s)=(t_s+\lambda_s\cos k_s) $ with $t_s$ and $\lambda_s$ being model parameters, and $h_s$ is the 1D extended SSH model introduced in Sec.~\ref{II}, respecting the chiral symmetry $C_s=i\gamma_{sa}^{(d)}\gamma_{sb}^{(d)}$. The $2^d\times 2^d$ matrices $\{\gamma_{1a,\cdots,da,1b,\cdots,db}^{(d)}\}$ form the Clifford algebra $\{\gamma_{j}^{(d)},\gamma_{j^{\prime}}^{(d)}\}=2\delta_{jj^{\prime}}$ for $j,j^{\prime}\in\{1a,\cdots,da,1b,\cdots,db\}$. The energy spectrum of $\mathcal{H}_{d}$ is
\beqn
E=\pm\sqrt{\sum\limits_{s=1}^{d}E_s^2}, \quad E_s= \sqrt{M_s^2(k_s)+(\lambda_s\sin k_s)^2},
\label{III2}
\eeqn
where $\pm E_s$ is the energy spectrum of $h_s$. Therefore, the bulk energy spectrum of $\mathcal{H}_{d}$ is fully gapped unless $|t_s|=|\lambda_s|$ for $s=1,\cdots,d$. Hereafter, we take $|t_s| <|\lambda_s|$ unless otherwise stated.

For the cases $d=1,2,3,4$, Hamiltonians $\mathcal{H}_d$ obtained by considering the concrete Gamma matrices representation have been previously studied \cite{Benalcazar2017a,Benalcazar2017,Chen2019b,Zhang2020e}. The bulk quadrupole ($d=2$), octupole ($d=3$), and hexadecapole ($d=4$) moments were identified by calculating the nested Wilson loop invariants.  In Eq.~\eqref{III1}, we generalize the BBH model to arbitrary dimensions from the perspective of Gamma matrices.

%When $d=1$, 1D Hamiltonian $\mathcal{H}_1$ behaves as the two-band SSH model, which hosts bulk dipole moment. When $d=2$ and $d=3$, $\mathcal{H}_2$ and $\mathcal{H}_3$ are the BBH model in 2D and 3D, corresponding to topological quadrupole and octupole insulators \cite{Benalcazar2017a,Benalcazar2017}, respectively.When $d=4$, $\mathcal{H}_4$ describes the topological hexadecapole insulators \cite{Chen2019b,Zhang2020e}.

%Based on these observations, we make a conjecture that $\mathcal{H}_d$ describes TMIs in an arbitrary $d$D system. We demonstrate this conjecture in the following.

\subsection{Corner zero-energy states}
\label{IIIB}
The BBH model hosts corner zero-energy states (CZESs), which are usually verified by numerically diagonalizing the tight-binding model Hamiltonian. For example, Hamiltonians $\mathcal{H}_2$, $\mathcal{H}_3$, and $\mathcal{H}_4$ have been shown to host $4$, 8, and 16 CZESs in 2D, 3D, and $4$D space, respectively \cite{Benalcazar2017a,Zhang2020e}. Here, we show that $\mathcal{H}_d$ generally hosts $2^d$ CZESs ($d\geq 2$) and construct the corresponding analytical wave function.

The construction of CZESs depends on the EZESs analytical solution of $h_s$ along each direction.
To construct the CZESs, we expand $\mathcal{H}_d(\bm k_d)$ at $\bm k_d=0$ to the second order of $\bm k_d$ and replace $\bm k_d=-i\partial_{\bm r_d}$, where $\bm r_d=(r_1,\cdots, r_d)$ is the coordinate vector. Under this low-energy expansion, then the EZESs of $h_s$ can be written as (see Eq.~\eqref{II3})
\beqn
|X_{z_s}^{(s)}(r_s)\rangle=f_{z_s}^{(s)}(r_s)|\psi_{z_s}^{(s)}\rangle,
\label{IIIB1}
\eeqn
where spinor $|\psi_{z_s}^{(s)}\rangle$ is the eigenstate of $C_s$ and satisfies $C_s|\psi_{z_s}^{(s)}\rangle=z_s|\psi_{z_s}^{(s)}\rangle$, with eigenvalue $z_s=\pm 1$, and $f_{z_s}^{(s)}(r_s)$  takes the identical form as $f_z(r)$ in Eq.~\eqref{II3} with an additional dimension label $s$.

Remarkably, we find that the equation $\mathcal{H}_d(\bm r_d)\Psi(\bm r_d)=0$ has $2^d$ solutions
\beqn
|\Psi_{z_1,\cdots,z_d}(\bm r_d)\rangle=\prod_{s=1}^df_{z_s}^{(s)}(r_s)|\psi_{z_1,\cdots,z_d}\rangle.
\label{Co1}
\eeqn
Here spinor $|\psi_{z_1,\cdots,z_d}\rangle$ belongs to the $2^d$ common eigenstates of matrices $\{C_1,\cdots,C_d\}$, which commute with each other, and $|\psi_{z_1,\cdots,z_d}\rangle$ satisfies
\beqn
C_s|\psi_{z_1,\cdots,z_d}\rangle=z_s|\psi_{z_1,\cdots,z_d}\rangle.
\eeqn
To give a concrete example, without loss of generality, we choose
\beqn
\begin{aligned}
&\gamma_{sa}^{(d)}=\underbrace{\sigma_3\otimes\cdots\sigma_3}_{s-1}\otimes\sigma_2\otimes\underbrace{\sigma_0\cdots\sigma_0}_{d-s},\\
&\gamma_{sb}^{(d)}=\underbrace{\sigma_3\otimes\cdots\sigma_3}_{s-1}\otimes\sigma_1\otimes\underbrace{\sigma_0\cdots\sigma_0}_{d-s}\\
&C_s=i\gamma_{sa}^{(d)}\gamma_{sb}^{(d)}=\underbrace{\sigma_0\otimes\cdots\sigma_0}_{s-1}\otimes\sigma_3\otimes\underbrace{\sigma_0\cdots\sigma_0}_{d-s},
\label{III3}
\end{aligned}
\eeqn
so that the $2^d$ common eigenstates of matrices $\{C_1,\cdots,C_d\}$  take the form
\beqn
|\psi_{z_1,\cdots,z_d}\rangle=\varphi_{z_1}\otimes\cdots\otimes\varphi_{z_s}\otimes\cdots\otimes \varphi_{z_d},
\label{III4}
\eeqn
where $\varphi_{z_s}=(1,0)^T$ for $z_s=1$ and $\varphi_{z_s}=(0,1)^T$ for $z_s=-1$, and the $2^d$ eigenstates of $|\psi_{z_1,\cdots,z_d}\rangle$ form a complete set of basis in $2^d$ spinor space.

By combining Eqs.~\eqref{IIIB1} and \eqref{Co1}, it can be easily verified that
\beqn
h_s(r_s)|\Psi_{z_1,\cdots,z_d}(\bm r_d)\rangle=0, \quad s=1,\cdots, d,
\eeqn
which gives rise to $\mathcal{H}_d(\bm r_d)|\Psi_{z_1,\cdots,z_d}(\bm r_d)\rangle=0$. As a wave function that decays along all the directions, $|\Psi_{z_1,\cdots,z_d}(\bm r_d)\rangle$ is localized at the corner, and therefore $\mathcal{H}_d$ hosts $2^d$ CZESs described by  Eq.~\eqref{Co1}.
According to Eq.~\eqref{Co1}, the localized position of CZES $|\Psi_{z_1,\cdots,z_d}(\bm r_d)\rangle$ in real space is completely determined by its labelled eigenvalues $\{z_1,\cdots,z_d\}$. To be specific, when $z_s=-1$ $(1)$,  CZES $|\Psi_{z_1,\cdots,z_d}(\bm r_d)\rangle$ is localized close to the corner where $r_s=0$ $(L)$, with $L$ being the length of the system along $r_s$. As clearly shown in Eq.~\eqref{III4}, the $2^d$ common eigenstates of $\{C_1,\cdots,C_d\}$ are not degenerate with respect to the eigenvalues $\{z_1,\cdots,z_d\}$, thus each corner of the hyper cubic space has one CZES.

%the $2^d$ CZESs of $\mathcal{H}_d$ are localized at each corner in the $d$D hyper cubic space.

Since there are $(2d+1)$ anti-commuting Gamma matrices with dimension $2^d\times 2^d$ (see Appendix \ref{Appendix I}), $d$D Hamiltonian $\mathcal{H}_d$ respects chiral symmetry represented by $\mathcal{C}=\prod_{s=1}^{d}C_s$,
with $\{\mathcal{C},\mathcal{H}_d\}=0$. Therefore, the CZESs are also the eigenstate of bulk chiral symmetry $\mathcal{C}$ with the eigenvalue $z=\prod_{s=1}^{d}z_s$. With this property, a local perturbation $h_p$ preserving the chiral symmetry $\mathcal{C}$ with $\{h_p,\mathcal{C}\}=0$ can not remove the zero-energy state localized at a given corner because
\beqn
\begin{aligned}
&\langle\Psi_{z_1,\cdots,z_d}|h_p|\Psi_{z_1,\cdots,z_d}\rangle\\
&=z\langle\Psi_{z_1,\cdots,z_d}|\mathcal{C}h_p+h_p\mathcal{C}|\Psi_{z_1,\cdots,z_d}\rangle/2=0.
\end{aligned}
\eeqn
 This analysis is consistent with the fact that the CZESs can survive when adding some additional terms preseving the chiral symmetry to $\mathcal{H}_2$ \cite{Li2020b}.

% For a similar reason, it can be shown that the chiral symmetries $\{C_1,\cdots,C_d\}$ also protect the CZESs.

In the above discussion, we construct CZESs under the low-energy continuum model. We emphasize that the presented 
method of constructing CZESs also applies to the lattice model Hamiltonian. In Appendix~\ref{Appendix III}, we construct the CZESs solution in the second quantization formalism.

\section{Boundary Hamiltonians }
\label{IV}
Based on the analytical solution of CZESs, we perform a general boundary projection to derive the boundary Hamiltonian of $\mathcal{H}_d$,  which turns out to be the BBH models of lower dimension and reveals the dimensional hierarchy.

\subsection{Boundary projection}

We schematically describe the process of extracting boundary Hamiltonians of $\mathcal{H}_d$ in two steps. First, we solve the $m$D gapped boundary states of $\mathcal{H}_d$ analytically for $m=d-1,\cdots,1$. Second, we project bulk Hamiltonian $\mathcal{H}_d$ onto the subspace expanded by the obtained boundary states, which leads to the boundary Hamiltonians.

To derive the gapped boundary states wave function of $\mathcal{H}_d$ under appropriate boundary conditions, we decompose Hamiltonian $\mathcal{H}_d$ into real space and momentum space parts
\beqn
\mathcal{H}_d({\bm r}_n,{\bm k}_m)=\bar{\mathcal{H}}_n({\bm r}_n)+\bar{\mathcal{H}}_m({\bm k}_m),
\label{RM}
\eeqn
where $\bm r_n=(r_{m+1},\cdots,r_d)$ is the $n$D real space coordinate vector, $\bm k_m=(k_1,\cdots,k_m)$ is the $m$D momentum vector, and $d=n+m$. With  a similar CZESs construction presented in Sec.~\ref{IIIB}, it can be shown that $\bar{\mathcal{H}}_n({\bm r}_n)$ also hosts $2^d$ EZESs ($n=1$) or CZESs ($n \geq 2$), which  are the common eigenstates of matrices $\{{C}_{m+1},\cdots, {C}_d\}$.
 Note that matrices ${C}_{m+1,\cdots,d}$ always commute with $\bar{\mathcal{H}}_m({\bm k}_m)$, namely $[{C}_{m+1,\cdots,d},\bar{\mathcal{H}}({\bm k}_m)]=0$. Therefore, the boundary states of $\mathcal{H}_d({\bm r}_n,{\bm k}_m)$, represented by $|\Phi_{\alpha}({\bm r}_n,{\bm k}_m)\rangle$, are the common eigenstates of ${C}_{m+1,\cdots,d}$ and $\bar{\mathcal{H}}({\bm k}_m)$ with the requirements
\beqn
\begin{aligned}
&\bar{\mathcal{H}}({\bm r}_n)|\Phi_{\alpha}({\bm r}_n,{\bm k}_m)\rangle=0,\\
&\bar{\mathcal{H}}({\bm k}_m)|\Phi_{\alpha}({\bm r}_n,{\bm k}_m)\rangle=E({\bm k}_m)|\Phi_{\alpha}({\bm r}_n,{\bm k}_m)\rangle,
\end{aligned}
\label{e16}
\eeqn
which gives rise to
\beqn
\mathcal{H}_d({\bm r}_n,{\bm k}_m)|\Phi_{\alpha}({\bm r}_n,{\bm k}_m)\rangle=E({\bm k}_m)|\Phi_{\alpha}({\bm r}_n,{\bm k}_m)\rangle.
\eeqn
Distinguished from the gapless boundary states of topological insulators \cite{Hasan2010,Qi2011}, these boundary states are gapped and extended over the whole boundary Brillouin zone, and therefore can be exactly described by a lattice Hamiltonian.

The boundary Hamiltonian can be extracted by projecting bulk Hamiltonian $\mathcal{H}_d$ onto the subspace expanded by the zero-energy states of $\bar{\mathcal{H}}({\bm r}_n)$. 
To illustrate this process, we take $d=2$ and $d=3$ for examples in the following.

\begin{figure}
\centering
\includegraphics[width=3.2in]{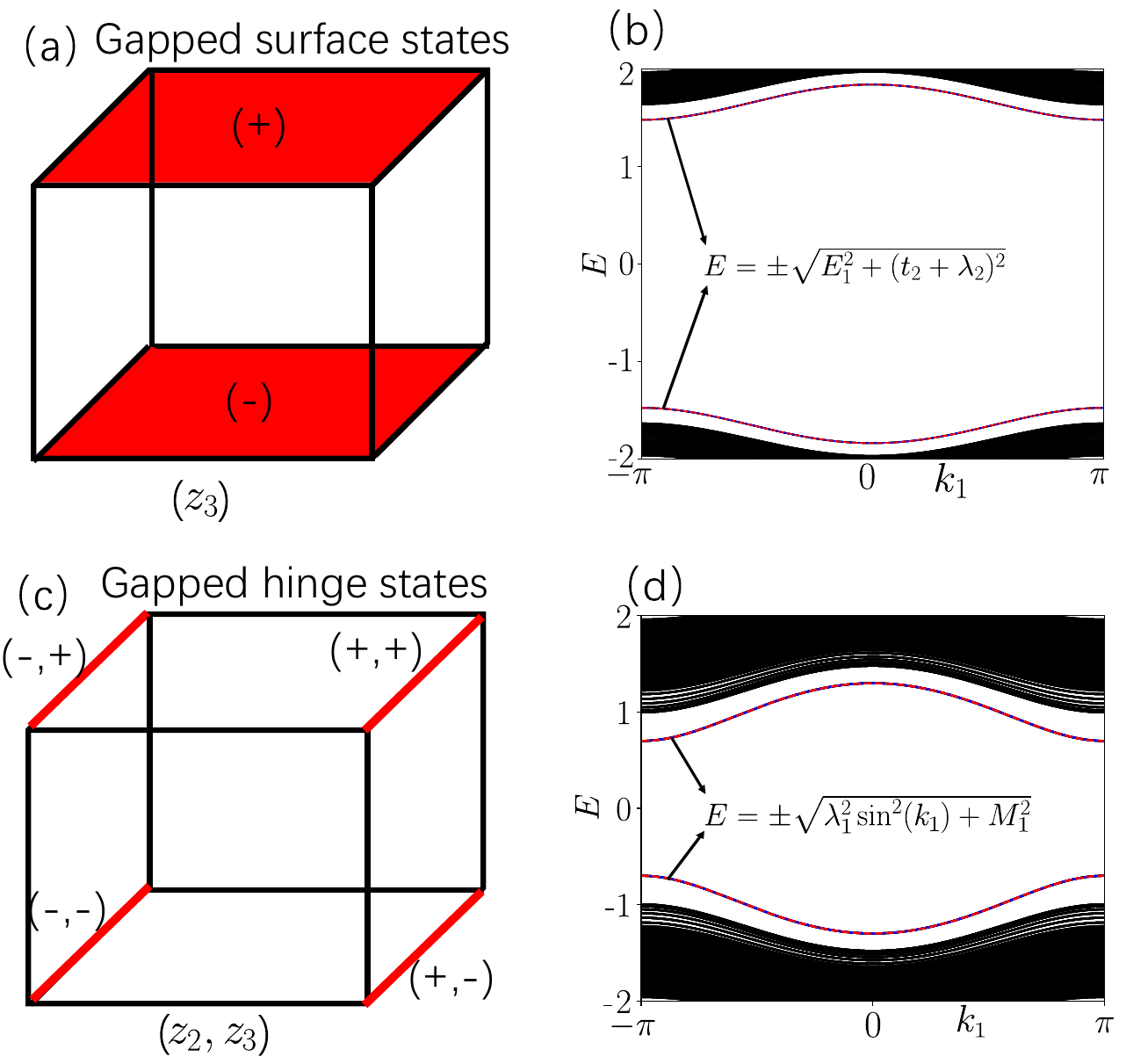}
\caption{(a) Schematic illustration of the gapped surface states labeled by the eigenvalue $(z_3)$ of $C_3$. (b) The energy spectrum along $k_1$ for a slab geometry of the system and we take $k_2=0$. (c)
Schematic illustration of the gapped hinge states labeled by  eigenvalues $(z_2,z_3)$ of $(C_2,C_3)$. (d) The energy spectrum along $k_1$ for a nanowire geometry of the system.
In (c) and (d), the energy spectrum of the in-gap red bands (red lines), corresponding to the gapped boundary states, can be accurately captured by the analytical results (dashed blue lines). For  (c) and (d), we take the model parameters as $t_s=0.3,\lambda_s=1$ with $s=1,2,3$.}
\label{SH}
\end{figure}

\subsection{ BBH models in $d=2$ and $d=3$}
\subsubsection{2D BBH model}
When $d=2$, $\mathcal{H}_d$ in Eq.~\eqref{III1} is specified by
\beqn
\begin{aligned}
\label{ad4}
&\mathcal{H}_{2}(\bm k_2)=\sum\limits_{s=1}^{2}h_s(k_s),\\
&h_s(k_s)=M_s(k_s)\gamma_{sa}^{(2)}+\lambda_s\sin k_s\gamma_{sb}^{(2)},
\end{aligned}
\eeqn
which yields $C_s=i\gamma_{sa}^{(2)}\gamma_{sb}^{(2)}$. When taking open and periodic boundary conditions, respectively,  along $s=2$ and $1$ directions, $\mathcal{H}_2(r_2,k_1)$ hosts gapped edge states \cite{Benalcazar2017a,Luo2022}. Since the EZESs of $h_2$ are the eigenstates of $C_2$ and $[C_2,h_1]=0$,
the gapped edge states of $\mathcal{H}_2(r_2,k_1)$
are the common eigenstates of $C_2$ and $h_1$ \cite{Khalaf2018,Khalaf2018a,Roberts2020}
\beqn
\begin{aligned}
&|\Phi_{z_2}(r_2,k_1)\rangle=f_{z_2}^{(2)}(r_2)P_{z_2}^{(2)}|\phi(k_1)\rangle,\\
&h_1(k_1)|\phi(k_1)\rangle=\pm E_1(k_1)|\phi(k_1)\rangle,
\end{aligned}
\eeqn
where edge projection operator is $P_{z_2}^{(2)}=(1+z_2C_2)/2$, and states $|\Phi_{-}(r_2,k_1)\rangle$ and $|\Phi_{+}(r_2,k_1)\rangle$ are localized close to edge $r_2=0$ and $r_2=L$, respectively. For a generic case, $|\phi(k_1)\rangle$ is the superposition of eigenstates of $C_2$ with eigenvalues of $1$ and $-1$. After the projection, the obtained state $|\Phi_{z_2}(r_2,k_1)\rangle$ is the eigenstate of $C_2$ with eigenvalue $z_2$.
%\beqn
%|\phi(k_1)\rangle=e_1|\phi(k_1)\rangle_{-}+e_2|\phi(k_1)\rangle_{+}
%\eeqn
%where $C_2|\phi(k_1)\rangle_{z_2}=z_2|\phi(k_1)\rangle_{z_2}$.
%Since $[C_2,h_1]=0$, the eigenstates of $h_1$ can be labeled by the eigenvalue of $C_2$, namely $z_2$. According to Eq.~\eqref{e16}, the gapped edge states of $\mathcal{H}_2(r_2,k_1)$ are the common eigenstates of $C_2$ and $h_1$ 

%\beqn
%\begin{aligned}
%&\Phi_{z_2}(r_2,k_1)=f_{z_2}^{(2)}(r_2)|\phi(k_1)\rangle_{z_2},\\
%&h_1(k_1)|\phi(k_1)\rangle_{z_2}=E_1(k_1)|\phi(k_1)\rangle_{z_2},
%\end{aligned}
%\eeqn
%where $|\phi(k_1)\rangle_{z_2}$ satisfies $C_2|\phi(k_1)\rangle_{z_2}=z_2|\phi(k_1)\rangle_{z_2}$.

The edge Hamiltonian can be obtained by projecting $\mathcal{H}_2$ onto the subspace defined by $P_{z_2}^{(2)}$. For example, the edge Hamiltonian for edge $r_2=0$ can be written as
\beqn
\begin{aligned}
\tilde{h}_1(k_1)&=P_{-}^{(2)}\mathcal{H}_2P_{-}^{(2)}\\
&=M_1(k_1)\tilde{\gamma}_{1a}^{(2)}+\lambda_1\sin k_1\tilde{\gamma}_{1b}^{(2)},
\end{aligned}
\eeqn
where $\tilde{\gamma}_{1a,1b}^{(2)}=P_{-}^{(2)}\gamma_{1a,1b}^{(2)}P_{-}^{(2)}$. As $[P_{-}^{(2)},h_2]=0$ and $(P_{-}^{(2)})^2=P_{-}^{(2)}$, we have $\{\tilde{\gamma}_{1a}^{(2)},\tilde{\gamma}_{1b}^{(2)} \}=0 $ and $(\tilde{\gamma}_{1a,1b}^{(2)})^2=P_{-}^{(2)}$. After projecting onto the non-zero block of $P_{-}^{(2)}$ \cite{Chiu2013,Khalaf2018}, $\tilde{\gamma}_{1a}^{(2)}$ and $\tilde{\gamma}_{1b}^{(2)}$ form the Clifford algebra and the $2\times 2$ edge Hamiltonian $\tilde{h}_1(k_1)$ behaves as a two-band SSH model. With the same physics, the other edges also behave as the two-band SSH model.

%which implies that $\{\tilde{\gamma}_{1a}^{(2)},\tilde{\gamma}_{1b}^{(2)} \}$ form the Clifford algebra once projected onto the non-zero block of $P_{-}^{(2)}$ \cite{Chiu2013,Khalaf2018}. Projecting onto the nonzero block of $P_{-}^{(2)}$,  To explicitly show this process, we follow the Gamma matrix representation given by Eq.~\eqref{III3}. In this case, we have
%\beqn
%C_2=\sigma_z\sigma_0,h_1=M_1\sigma_z\sigma_x+\lambbda_1\sink_1
%\eeqn

\subsubsection{3D BBH model}
When $d=3$, $\mathcal{H}_d$ is given by
\beqn
\begin{aligned}
&\mathcal{H}_{3}(\bm k_3)=\sum\limits_{s=1}^{3}h_s(k_s),\\
&h_s(k_s)=M_s(k_s)\gamma_{sa}^{(3)}+\lambda_s\sin k_s\gamma_{sb}^{(3)},
\end{aligned}
\eeqn
which gives $C_{s}=i\gamma_{sa}^{(3)}\gamma_{sb}^{(3)}$.
Under the periodic and open boundary conditions, respectively, along $s=1,2$ and $s=3$ directions, we decompose $\mathcal{H}_{3}$ as $\mathcal{H}_{3}(r_3,k_1,k_2)=h_3(r_3)+h_{12}(k_1,k_2)$, with $h_{12}=h_1+h_2$. Similarly, the gapped surface states of $\mathcal{H}_{3}(r_3,k_1,k_2)$, schematically illustrated in Fig.~\ref{SH}(a), are the common eigenstates of $C_3$ and $h_{12}$
\beqn
\begin{aligned}
&|\Phi_{z_3}(r_3,k_1,k_2)\rangle=f_{z_3}^{(3)}(r_3)P_{z_3}^{(3)}|\phi(k_1,k_2)\rangle,\\
&h_{12}(k_1,k_2)|\phi(k_1,k_2)\rangle=E(k_1,k_2)|\phi(k_1,k_2)\rangle,
\end{aligned}
\eeqn
where surface projection operator is $P_{z_3}^{(3)}=(1+z_3C_3)/2$, and the surface energy spectrum is $E=\pm\sqrt{E_1^2+E_2^2}$, as shown in Fig.~\ref{SH}(b).

Similar to the $d=2$ case, the surface Hamiltonian for surface $r_3=0$ can be written as
\beqn
\begin{aligned}
\tilde{h}_{12}(k_1,k_2)&=P_{-}^{(3)}\mathcal{H}_3P_{-}^{(3)}\\
&=\sum\limits_{s=1}^2M_s(k_s)\tilde{\gamma}_{sa}^{(3)}+\lambda_s\sin k_s\tilde{\gamma}_{sb}^s,
\end{aligned}
\eeqn
where we have defined $\tilde{\gamma}_{sa,sb}^{(3)}=P_{-}^{(3)}\gamma_{sa,sb}^{(3)}P_{-}^{(3)}$. Projecting onto the non-zero block of $P_{-}^{3}$, the $4\times 4$ surface Hamiltonian $\tilde{h}_{12}(k_1,k_2)$ takes completely identical form as the 2D BBH model Hamiltonian $\mathcal{H}_2$.

%Following the same reasons as $d=2$ case, we can 

To further extract the hinge Hamiltonian, we take open (periodic) boundary conditions, respectively,  along $s=2,3$ ($s=1$) directions, under which we decompose $\mathcal{H}_{3}$ as $\mathcal{H}_{3}(r_2,r_3,k_1)=h_{23}(r_2,r_3)+h_1(k_1)$, with $h_{23}=h_2+h_3$. The gapped hinge states of $\mathcal{H}_{3}(r_2,r_3,k_1)$, schematically illustrated in Fig.~\ref{SH}(c), are the common eigenstates of $C_{2,3}$ and $h_{1}$
\beqn
\begin{aligned}
&|\Phi_{z_2,z_3}(r_2,r_3,k_1)\rangle=\prod_{s=2}^3f_{z_s}^{(s)}P_{z_s}^{(s)}|\phi(k_1)\rangle,\\
&h_1(k_1)|\phi(k_1)\rangle=\pm E_1(k_1)|\phi(k_1)\rangle,
\end{aligned}
\eeqn
where $P_{z_s}^{(s)}=(1+z_sC_s)/2$ and the gapped hinge energy spectrum is $E=\pm E_1$, as shown in Fig.~\ref{SH}(d). When $(z_2,z_3)=\{(-,-),(-,+),(+,-),(+,+)\}$,  gapped hinge state $|\Phi_{z_2,z_3}(r_2,r_3,k_1)\rangle$ is localized close to the hinges $(r_2,r_3)=\{(0,0),(0,L),(L,0),(L,L)\}$, respectively.

The hinge Hamiltonian for  hinge $(r_2,r_3)=(0,0)$ can be written as
\beqn
\begin{aligned}
\tilde{h}_1(k_1)&=P_{--}\mathcal{H}_3P_{--}\\
&=M_1(k_1)\tilde{\gamma}_{1a}^{(3)}+\lambda_1\sin k_1\tilde{\gamma}_{1b}^{(3)},
\end{aligned}
\eeqn
where  hinge projection operator is $P_{z_2z_3}=\prod_{s=2}^3P_{z_s}^{(s)}$ and $\tilde{\gamma}_{1a,1b}^{(3)}=P_{--}\gamma_{1a,1b}^{(3)}P_{--}$. Distinguished from $P_{-}^{(3)}$, the non-zero block of $P_{--}$ is $2\times 2$ as $P_{--}$ is the product of two projection operators. Thus when projecting onto the non-zero block of $P_{--}$,  $\tilde{h}_1(k_1)$ takes the exact form of a two-band SSH model.

%Both the surface quadrupole and hinge dipole moments manifest bulk octupole moment.

\begin{table}[htb]
\centering
\setlength\tabcolsep{1.6pt}
\renewcommand{\arraystretch}{1.5}
\caption{ The hierarchy of the BBH model in arbitrary dimensions. The second row denotes the dimension of the bulk or boundary of the system.
The third row lists the bulk Hamiltonian $\mathcal{H}_{d}$ and boundary Hamiltonians $\mathcal{\tilde{H}}_{1,\cdots,d-1}$ with the space dimension in the second row. The fourth row represents the matrix dimension for the Hamiltonians in the third row. The fifth row denotes the number of the CZESs or EZESs for the Hamiltonians in the third row. The sixth row denotes the used boundary projectors to extract the boundary Hamiltonians in the second row. The seventh row denotes the number of configurations of the boundary projections listed in the sixth row.}
\begin{tabular}{|c|c|c|c|c|c|c|}
\hline
{} & bulk &\multicolumn{5}{c|}{boundary} \\
\hline
dimensions & $d$ & $d-1$ & $d-2$ & $\cdots$ & $2$ & $1$ \\
\hline
Hamiltonians & $\mathcal{H}_d$ & $\mathcal{\tilde{H}}_{d-1}$ & $\mathcal{\tilde{H}}_{d-2}$ &$\cdots$ & $\mathcal{\tilde{H}}_{2}$ & $\mathcal{\tilde{H}}_{1}$\\
\hline
matrices dimension& $2^d$ & $2^{d-1}$ & $2^{d-2}$ & $\cdots$ & $2^2$ & $2^1$ \\
\hline
\makecell{number of CZESs \\ or EZESs }& $2^d$ & $2^{d-1}$ & $2^{d-2}$ & $\cdots$ & $2^2$ & $2^1$ \\
\hline
boundary projectors &~&$P_{z_d}$&$P_{z_{d-1}z_d}$&$\cdots$ &$P_{ z_3\cdots z_d}$&$P_{ z_2 \cdots z_d}$\\
\hline
\makecell{number of \\ configurations} &~ & $2^{1}$ & $2^{2}$ & $\cdots$ & $2^{d-2}$ & $2^{d-1}$ \\
\hline
\end{tabular}
\label{tab}
\end{table}

\subsection{Generalization to arbitrary dimension }
We now show that the $m$D boundary Hamiltonians of $\mathcal{H}_{d}$ take completely identical form as the $m$D BBH model Hamiltonian $\mathcal{H}_m$ in the following.

%which exhibits the boundary localized multipole moment and indicates the existence of bulk multipole moment.  

%Therefore, $\mathcal{H}_{d}$ describes TMIs as the existence of boundary-localized multipole moments of lower dimension.

Following Eq.~\eqref{RM}, $\mathcal{\bar{H}}_n$ hosts $2^d$ CZESs which are the common eigenstates of $\{
C_{m+1},\cdots,C_d\}$ and can be written as
\beqn
&|\Psi_{z_{m+1},\cdots, z_d}({\bm r}_{n})\rangle=\prod\limits_{q=m+1}^{d}f_{z_q}^{(q)}(r_q)|\psi_{z_{m+1},\cdots ,z_d}\rangle.
\eeqn
The zero-energy subspace expanded by the CZESs of $\mathcal{\bar{H}}_n({\bm r}_n)$ can be defined by the boundary projection operator
\beqn
P_{z_{m+1}\cdots z_d}=\prod_{q=m+1}^{d}(1+z_qC_q)/2.
\eeqn
The gapped boundary states of $\mathcal{H}_d({\bm r}_n,{\bm k}_m)$ are the common eigenstates of $C_{m+1,\cdots,d}$ and $\mathcal{\bar{H}}_m$
\beqn
&&|\Phi_{z_{m+1}\cdots z_d}({\bm {r}}_{n},{\bm {k}}_m)\rangle=\prod_{q=m+1}^{d}f_{z_q}^{(q)}(r_q)P_{z_{m+1}\cdots z_d}|\phi({\bm {k}}_m)\rangle,\nonumber\\
&&\bar{\mathcal{H}}_m({\bm k}_m)|\phi({\bm {k}}_m)\rangle= E({\bm k}_m)|\phi({\bm {k}}_m)\rangle,
\eeqn
where $E(\bm k_m)=\pm \sqrt{\sum_{s=1}^{m} E_s^2(k_s)}$.
For different eigenvalues configurations of $(z_{m+1}\cdots z_d)$, state $|\Phi_{z_{m+1}\cdots z_d}(\bm {r}_n,\bm {k}_{m})\rangle$ is localized close to different boundaries.

Boundary Hamiltonians can be derived by projecting $\mathcal{H}_d$ onto the zero-energy subspace defined by $P_{z_{m+1}\cdots z_d}$.
For example, the boundary Hamiltonian for boundary $r_{m+1,\cdots,d}=0$ can be extracted as
\beqn
\begin{aligned}
\tilde{\mathcal{H}}_m(\bm k_m)&=P_{-\cdots-}\mathcal{H}_dP_{-\cdots-}\\
&=\sum\limits_{l=1}^{m}M_l(k_l)\tilde{\gamma}_{la}^{(d)}+\lambda_l\sin(k_l)\tilde{\gamma}_{lb}^{(d)},
\end{aligned}
\eeqn
where $\tilde{\gamma}_{la, lb}^{(d)}=P_{-\cdots-}\gamma_{la,lb}^{(d)}P_{-\cdots-}$.
As $[P_{-\cdots-}, \gamma_{la,lb}^{(d)}]=0$, we have $\{\tilde{\gamma}_{u}^{(d)},\tilde{\gamma}_{u^{\prime}}^{(d)}\}=2\delta_{uu^{\prime}}P_{-\cdots-}$ for $u,u^{\prime}=1a,\cdots,ma,1b,\cdots,mb$.
After projecting onto the non-zero block of $P_{-\cdots-}$, corresponding to the eigenvalues $z_{m+1,\cdots,d}=-1$, $\{\tilde{\gamma}_{1a,\cdots,ma}^{(d)},\tilde{\gamma}_{1b,\cdots,mb}^{(d)} \}$ form Clifford algebra and boundary Hamiltonian $\tilde{\mathcal{H}}_m$ takes the exact form of the $m$D BBH model $\mathcal{H}_{m}$.
  We note that the derived boundary Hamiltonian $\tilde{H}_m$ is related to the higher-order band inversion surfaces given by a special region in the
Brillouin zone where $M_s(k_s)=0$ for $s=1,\cdots,d$ as studied in \cite{Jia2022}

%When $m=1,2,3$, and $4$, $\tilde{\mathcal{H}}_m$ take the exact form of the two band SSH model, giving rise to the boundary-localized dipole, quadrupole, octupole, and hexadecapole moments, respectively, indicating the existence of bulk multipole moment.

%given by a special region in the Brillouin zone with vanishing pseudo-spin components of the Hamiltonian as studied in the work \cite{Jia2022}. The existence of band-inversion surfaces implies the existence of boundary states, namely dual between mass domain walls on real-space boundaries and higher-order band-inversion surfaces \cite{Jia2022}. With this correspondence, the work \cite{Jia2022} proposed a momentum-space topological characterization of a class of higher-order topological phases. Similar topological characterization could be applied to the BBH model with some revisions. 

The multidimensional boundary states  of the BBH model reflect the dimensional hierarchy, summarized in Table~\ref{tab}. Meanwhile, there are rich topological transitions for the BBH model $\mathcal{H}_d$. When $|t_s|=|\lambda_s|$ for $s=1,\cdots,d$, the energy gap of $\mathcal{H}_d$ closes, associated with bulk topological phase transition. When $d-n$ pairs of parameters $(t_s,\lambda_s)$ satisfy $|t_s|=|\lambda_s|$ and the other $n$ pairs of parameters $(t_{s^{\prime}},\lambda_{s^{\prime}})$ satisfy $|t_{s^{\prime}}|<|\lambda_{s^{\prime}}|$,
the energy gap of certain $(d-n)$D boundary closes, corresponding to a boundary topological phase transition. Both the bulk and boundary topological phase transitions can remove the corner states.

\section{Discussion and Conclusion}
\label{V}

 In summary, we generalize the BBH model to arbitrary dimensions and reveal their dimensional hierarchy. For the BBH in arbitrary dimensions, the analytical solution of CZESs can be directly constructed. Based on the analytical solution of CZESs, we develop a general boundary projection method to exact the boundary Hamiltonians, which turns out to be the BBH model of lower dimension and unveils the inherent hierarchy of the BBH model.

For the 2D and 3D BBH models, the nontrivial topology can be characterized by the invariants such as nested Wilson loop, multipole moment calculated based on the many-body wave function \cite{Wheeler2019,Kang2019}, and recently proposed chiral multipole number \cite{Benalcazar2022}. Although these invariants can be generalized to higher-dimensional systems,  it is generally hard to numerically calculate these topological invariants in a system with a dimension larger than three. Since the $d$D BBH model is constructed by the combination of the 1D extended SSH model
along different directions, the nontrivial topology of the BBH models in arbitrary dimension can be naturally characterized by the topological invariant $\nu=\prod_{i=1}^{d}\nu_i$, where $\nu_i=1$ ($\nu_i=0$) corresponding to the presence ( absence) of the end zero-energy states for $h_i$. We note that a similar topological invariant was defined to characterize a class of higher-order topological phases in arbitrary dimensions with arbitrary orders \cite{Jia2022}.
%This topological invariant can capture the nontrivial topology of the BBH model in arbitrary dimensions.
%A similar topological invariant recently was proposed to characterize a class of higher-order topological phase in arbitrary dimensions with arbitrary orders,

In $\mathcal{H}_d$, although we use the Gamma matrices that anti-commute with each other, this condition can be relaxed to obtain CZESs.  In the process of constructing CZESs, we can find that as long as
matrices $\{C_1,\cdots,C_d\}$ commute with each other, $2^d$ CZESs can be constructed in the same way for Hamiltonian $\mathcal{H}_d$, where the 2D case was studied in the work \cite{Luo2022}. Therefore, our construction method of CZESs applies to a large class of HOTPs which are stacked by the 1D SSH model along different directions.
On the other hand,  we only consider the nearest neighboring hopping couplings in $\mathcal{H}_d$. It was shown that the long-range hopping in the BBH model can give rise to multiple corner zero-energy states at each corner \cite{Benalcazar2022}. We emphasize that our theory of constructing corner zero-energy states by the end zero-energy states of different directions is still valid when considering long-range hopping couplings as long as the Bloch Hamiltonian can be separated as $\mathcal{H}_d(\bm k_d)=\sum_{s=1}^{d}h_s(k_s)$.

The topological property of Hamiltonian $\mathcal{H}_d$ does not depend on the chosen
physical basis. If $\mathcal{H}_d$ is written
in the superconducting Bogoliubov-de Gennes basis,
the generalized BBH models describe
higher-order topological superconductors featuring Majorana corner states, which could be used to perform topological quantum computation \cite{Nayak2008}. We note that the 2D BBH model in the superconducting system has been studied in the literature \cite{Wang2018a,Tiwari2020a}

Recent advances in synthetic dimensions provide a feasible way to investigate higher-dimensional topological physics in lower dimensionality by means of additional parameter space. For example, the 4D quantum Hall effect has been realized in photonic crystals  \cite{Zilberberg2018} and cold atom systems \cite{Lohse2018} by utilizing synthetic dimensions. In addition, the higher-dimensional system in electric circuits can be flexibly constructed by connecting nonlocal sites, which provides a good platform to study the higher-dimensional physics \cite{ME2019,Lih2019,yu2020}. Thus, the higher-dimensional BBH model could be simulated in these artificial systems.

\section{ACKNOWLEDGMENTS}
We thank Xin Liu and Xiao-Hong Pan for their helpful discussions. This work is supported by Key Research and Development Program of Hubei Province (Grant No. 2022BAA017) and start-up funding of Wuhan University.

\appendix

\section{Clifford algebra and Gamma matrices}
\label{Appendix I}
For the complex Clifford algebra $Cl_{2d}$, there are $2d$ generators, labelled as $e_{j}$ for $j=1,\cdots,2d$, which satisfy the anti-commutation relation $\{e_{j},e_{j^{\prime}}\}=2\delta_{jj^{\prime}}$. Another operator anti-commutes with the $2d$ generators and can be written as
\beqn
e_{2d+1}=i^ne_1e_2,\cdots,e_{2d}, \{e_{2d+1}, e_j\}=0,e_{2d+1}^2=1.\nonumber\\
\eeqn
The $(2d+1)$ anti-commuting operators $e_{1,\cdots,2d+1}$ can be represented in terms of anti-commuting $2^d\times 2^d$ Gamma matrices $\gamma^{(d)}_{1,2,\cdots,2d+1}$, which are the direct product of $d$ sets of Pauli matrices and $2\times 2$ identify matrix and satisfy $(\gamma_j^{(d)})^{2}=1,(\gamma_j^{(d)})^{\dagger}=\gamma_j^{(d)}$.

When $d$=1, $\gamma_{1,2,3}^{(1)}$ correspond to the three Pauli matrices $\sigma_{1,2,3}$. When $d$=2, $\gamma_{1,2,3,4,5}^{(2)}$ correspond to five anti-commuting Dirac matrices, which can be chosen as
\beqn
\begin{aligned}
&\gamma_{1,2,3}^{(2)}=\sigma_3\otimes \sigma_{1,2,3}, \gamma_4^{(2)}=\sigma_1\otimes \sigma_{0},\\
&\gamma_5^{(2)}=-\gamma_1^{(2)}\gamma_2^{(2)}\gamma_3^{(2)}\gamma_4^{(2)}=\sigma_2\otimes \sigma_0.
\end{aligned}
\eeqn
The anti-commuting $2^d\times 2^d$ Gamma matrices $\gamma^{(2d)}_{1,2,\cdots,2d+1}$ can be generically obtained according to the iteration relation
\beqn
\begin{aligned}
&\gamma_{1,2,\cdots,2d-1}^{(d)}=\sigma_3\otimes \gamma_{1,2,\cdots,2d-1}^{(d-1)},\\
& \gamma_{2d}^{(d)}=\sigma_1\otimes I, \gamma_{2d+1}^{(d)}=\sigma_2\otimes I,
\end{aligned}
\eeqn
where $ I$ denotes the $2^{d-1}\times 2^{d-1}$ identify matrix.

\section{1D extended SSH model}
\label{Appendix II}
We present the calculation of the winding number and EZESs wave function for the 1D extended SSH model. Following the extended SSH model Hamiltonian in Eq.~\eqref{II1},  the energy spectrum of $h$ is $\pm E_0=\pm\sqrt{(t+\lambda\cos k)^2+(\lambda\sin k)^2}$,  then $h$ can be normalized as
\beqn
\begin{aligned}
\bar{h}&=h/E_0\\
&=\cos\varphi\gamma_{a}^{(d)}+\sin\varphi\gamma_{b}^{(d)},
\end{aligned}
\eeqn
with $\cos\varphi=(t+\lambda\cos k)/E_0$ and  $\sin\varphi=\lambda\sin k/E_0$. The winding number contributed by the occupied states of $h(k)$ is \cite{Zhang2018a}
\beqn
\begin{aligned}
{\nu}&=\frac{1}{4i\pi }\int_{-\pi}^{\pi}\text{Tr}[C\bar{h}\partial_k\bar{h}]dk\\
&=\frac{1}{4i\pi }\int_{-\pi}^{\pi}\text{Tr}[(\cos\varphi\partial_{k}\cos\varphi+\sin\varphi\partial_{k}\sin\varphi)C+\\
&\quad (\cos\varphi\partial_{k}\sin\varphi-\sin\varphi\partial_{k}\cos\varphi)C\gamma_{1}^{(d)}\gamma_{2}^{(d)}]dk\\
&=\frac{2^d}{4\pi }\int_{-\pi}^{\pi}(\cos\varphi\partial_{k}\sin\varphi-\sin\varphi\partial_{k}\cos\varphi)dk\\
&=\frac{2^d}{4\pi }\int_{-\pi}^{\pi}\partial_{k}\varphi dk.
\label{wind}
\end{aligned}
\eeqn
Here, we have used the traceless property of matrix $C$. When $|t|<|\lambda|$, the integration in Eq.~\eqref{wind} gives ${\nu}=2^{(d-1)}$, otherwise ${\nu}=0$.

The winding number ${\nu}=2^{(d-1)}$ implies that $h$ hosts $2^{(d-1)}$ EZESs at each end under the open boundary condition. Expanding $h$ at $k=0$ (supposing that $E_0$ takes minimum value at $k=0$ ) to the second order of $k$ and replacing $k\rightarrow -i\partial_{r}$, we have
\beqn
h({r})=(m+\lambda/2\partial_{r}^2)\gamma_{a}^{(d)}-i\lambda\partial_{r}\gamma_{b}^{(d)},
\eeqn
with $m=t+\lambda$. Considering the semi-infinite system ($r>0$), the EZESs localized close to $r=0$ can be derived by solving the equation
\beqn
(m+\lambda/2\partial_{r}^2)\gamma_{a}^{(d)}|X_{\alpha}(r)\rangle-i\lambda\partial_{r}\gamma_{b}^{(d)}|X_{\alpha}(r)\rangle=0.
\label{S1}
\eeqn
Multiplying both sides by $\gamma_{1}^{(d)}$ in Eq.~\eqref{S1} yields
\beqn
(m+\lambda/2\partial_{r}^2)|X_{\alpha}(r)\rangle=\lambda\partial_{r}C|X_{\alpha}(r)\rangle.
\label{eq}
\eeqn
Therefore, state $|X_{\alpha}(r)\rangle$ is the eigenstate of chiral symmetry $C$ and is labeled by the chiral symmetry eigenvalue $z=\pm 1$. We set the trial wave function $|X_{z}(r)\rangle=e^{\xi_{z} r}|\psi_{z}\rangle$, where $\xi_{z}$ is a complex number, and spinor $|\psi_{z}\rangle$ satisfies $C|\psi_{z}\rangle=z|\psi_{z}\rangle$. Inserting this ansatz solution into Eq.~\eqref{eq}, we have
\beqn
\lambda/2\xi_{z}^2-z\lambda\xi+m=0,
\eeqn
which gives two roots $ \xi_{z}^{1,2}=\frac{z\lambda \pm \sqrt{\lambda^2-2m\lambda}}{\lambda}$. When $|t|<|\lambda|$, the real part of $ \xi_{z}^{1,2}$ are negative and positive when $z=-1$ and $z=1$, respectively. Under the boundary conditions $|X_z(0)\rangle=|X_z(\infty)\rangle=0$, the wave function of the EZESs can be written as
\beqn
\begin{aligned}
&|X_{-}(r)\rangle=f_{-}(r)|\psi_{-}\rangle,\\
& f_{-}(r)=\mathcal{N}_{-}(e^{\xi_{-}^{1}r}-e^{\xi_{-}^{2}r}),
\end{aligned}
\eeqn
where $\mathcal{N}_{-}$ is the normalization factor. Similarly, if we consider the semi-system $r<0$, the EZESs localized close to $r=0$ is the eigenstate of $C$ with eigenvalue $z=1$. Therefore, for a finite system with length $L$, the EZESs localized close to the end $r=0$ and $r=L$ are the eigenstates of $C$, with eigenvalue $z=-1$ and $z=1$, respectively. Analogously, the wave function of CZESs localized at $r=L$ can be written as
\beqn
\begin{aligned}
&|X_{+}(r)\rangle=f_{+}(r)|\psi_{-}\rangle,\\
&f_{+}(r)=\mathcal{N}_{+}(e^{\xi_{+}^1(r-L)}-e^{\xi_{+}^2(r-L)}),
\end{aligned}
\eeqn
where $\mathcal{N}_{+}$ is the normalization factor.

\section{The CZESs solution in the second quantization formalism}
\label{Appendix III}
We show that the constructed wave function in Eq.~\eqref{Co1} can also be obtained using the second quantization formalism.
We rewrite the Hamiltonian in Eq.~\eqref{III1} as
\beqn
\begin{aligned}
&\mathcal{H}^{(d)}=\sum_{\alpha,\beta,\bm k_d} \mathcal{H}_{\alpha,\beta}^{(d)}(\bm k_d) c_{\alpha,\bm k_d}^{\dagger}c_{\beta,\bm k_d},\\
&\mathcal{H}_{\alpha,\beta}^{(d)}(\bm k_d)=\sum_{s=1}^{d} h_{\alpha,\beta}^{(s)}(k_s),\\
&h^{(s)}(k_s)=M_s(k_s)\gamma_{sa}^{(d)}+\lambda_s\sin k_s\gamma_{sb}^{(d)},
\label{31}
\end{aligned}
\eeqn
where indexes $\alpha$ and $\beta$ denote the inner degree of freedom of Hamiltonian with $\alpha,\beta=1,\cdots,2^d$, $\bm k_d=(k_1,\cdots,k_d)$ is $d$D momentum vector, index $s$ denotes the different directions, and $c_{\alpha,\bm k_d}^{\dagger}$ ($c_{\beta,\bm k_d}$) denotes the creation (annihilation) operator at $\bm k_d$ with sublattice $\alpha$ ($\beta$). $h^{(s)}$ is the 1D extended SSH model and respects chiral symmetry $C_s=i\gamma_{sa}^{(d)}\gamma_{sb}^{(d)}$.

By performing Fourier transformation for $\mathcal{H}_d$, we obtain the tight-binding model Hamiltonian
\beqn
\begin{aligned}
&\mathcal{H}^{(d)}=\sum_{\alpha,\beta,\bm i_d,\bm i_d^{\prime}} \mathcal{H}_{\alpha,\beta}^{(d)}(\bm i_d,\bm i_d^{\prime})c_{\alpha,\bm i_d}^{\dagger}c_{\beta,\bm i_d^{\prime}},\\
&\mathcal{H}^{(d)}_{\alpha,\beta}(\bm i_d,\bm i_d^{\prime})=\sum_{s=1}^{d}h_{\alpha,\beta}^{(s)}(i_s,i_s^{\prime}),
\end{aligned}
\eeqn
where $\bm i_d=(i_1,\cdots,i_d)$ and $\bm i_d^{\prime}=(i_1^{\prime},\cdots,i_d^{\prime})$ denote the coordinate of lattice site in the $d$D system, $i_s$ and $i_s^{\prime}$ take the integer value, and
$c_{\alpha,\bm i_d}^{\dagger}$ ($c_{\beta,\bm i_d^{\prime}}$) is the creation (annihilation) operator at site $\bm i_d$ ($\bm i_d^{\prime}$) with sublattice $\alpha$ ($\beta$). 

In the second quantization formalism,
the $2^d$ EZESs of $h^{(s)}$ can be generically written as
\beqn
|X^{(s)}\rangle=\sum_{\bm{i_d},\eta}f^{(s)}(i_s)\chi_{\eta}c_{\eta,\bm i_d}^{\dagger}|0\rangle,
\eeqn
where $\bm i_d=(i_1,\cdots,i_d)$ denotes the lattice coordinate, $f^{(s)}(i_s)$ is the site-dependent wave function, spinor $\chi$ belongs to the $2^d$ common eigenstates of $\{C_1,\cdots,C_d\}$ with component $\chi_{\eta}$, and $|0\rangle$ denotes the vacuum state.

The Hamiltonian $h^{(s)}$ acts on the EZESs $|X^{(s)}\rangle$ as
\beqn
h^{(s)}|X^{(s)}\rangle&=&\sum_{\alpha,\beta,\eta,\bm i_d,\bm i_d^{\prime},\bm i_d^{\prime\prime}}h_{\alpha,\beta}^{(s)}(i_s,i_s^{\prime})c_{\alpha,\bm i_d}^{\dagger}c_{\beta,\bm i_d^{\prime}}f^{(s)}(i_s^{\prime\prime})\chi_{\eta}c_{\eta,\bm i_d^{\prime\prime}}^{\dagger}|0\rangle\nonumber\\
&=&\sum_{\alpha,\beta,\bm i_d,\bm i_d^{\prime}}h_{\alpha,\beta}^{(s)}(i_s,i_s^{\prime})f^{(s)}(i_s^{\prime})\chi_{\beta}c_{\alpha,\bm i_d}^{\dagger}|0\rangle,\nonumber\\
\label{c4}
\eeqn
which vanishes if
\beqn
\sum_{\beta,\bm i_d^{\prime}}h_{\alpha,\beta}^{(s)}(i_s,i_s^{\prime})f^{(s)}(i_s^{\prime})\chi_{\beta}=0.
\label{32}
\eeqn
For the Hamiltonian $
\mathcal{H}^{(d)}$, we construct the corner-localized state
\beqn
|\Psi\rangle=\sum_{\bm {i_d},\eta}\prod_{s=1}^{d}f^{(s)}(i_{s})\chi_{\eta}c_{\eta,\bm i_d}^{\dagger}|0\rangle.
\eeqn
Now we demonstrate that $\mathcal{H}^{(d)}|\Psi\rangle=0$.  Each term $h^{(s)}$, for $s=1,\cdots,d$,
acts on the corner-localized state $|\Psi\rangle$ as
\begin{widetext}
\beqn
h^{(s)}|\Psi\rangle &=&\sum_{\alpha,\beta,\eta,\bm i_d,\bm i_d^{\prime},\bm i_d^{\prime\prime}} h_{\alpha,\beta}^{(s)}(i_{s},i_{s}^{\prime})c_{\alpha,\bm i_d}^{\dagger}c_{\beta,\bm i_d^{\prime}}\prod_{s^{\prime}=1}^{d}f^{(s^{\prime})}(i_{s^{\prime}}^{\prime\prime})\chi_{\eta}c_{\eta,\bm i_d^{\prime\prime}}^{\dagger}|0\rangle\nonumber\nonumber\\
&=&\sum_{\alpha,\beta,\bm i_d,\bm i_d^{\prime}}h_{\alpha,\beta}^{(s)}(i_{s},i_{s}^{\prime})f^{(s)}(i_{s}^{\prime})\chi_{\beta}\prod_{s^{\prime}\neq s}f^{(s^{\prime})}(i_{s^{\prime}}^{\prime})c_{\alpha,\bm i_d}^{\dagger}|0\rangle.
\eeqn
\end{widetext}
Combining with Eq.~\eqref{32}, we have $h^{(s)}|\Psi_{\eta}\rangle=0$.
Therefore, the state $|\Psi\rangle$ is a CZES of $\mathcal{H}^{d}$.

\bibliography{reference1}

\end{document}